\documentclass{aa}
\usepackage[dvips]{graphicx}
\usepackage{natbib}
\usepackage{paralist}

\usepackage{subfigure}
\bibpunct{(}{)}{;}{a}{}{,}
\begin{document}

\title{Multi-wavelength observations and modelling of a canonical solar flare}
\author{Claire~L. Raftery\inst{1, 2}
  \and Peter~T. Gallagher\inst{1}
  \and Ryan~O. Milligan\inst{2} 
  \and James~A. Klimchuk\inst{2}}
  
\institute{Astrophysics Research Group, School of Physics, Trinity College Dublin, Dublin 2, Ireland.
  \and Solar Physics Laboratory (Code 671), Heliophysics Science Division, NASA Goddard Space Flight Centre, Greenbelt, MD 20771, U.S.A.}
  
\date{Received:, Accepted:}
\abstract
{}{To investigate the temporal evolution of temperature, emission measure, energy loss and velocity in a C-class solar flare from both an observational and theoretical perspective. 
}
{
The properties of the flare were derived by following the systematic cooling of the plasma through the response functions of a number of instruments -- the Reuven Ramaty High Energy Solar Spectroscopic Imager (RHESSI; $>$5~MK), GOES-12 (5--30~MK), the Transition Region and Coronal Explorer (TRACE 171~\AA; 1~MK) and the Coronal Diagnostic Spectrometer (CDS; $\sim$0.03--8~MK). These measurements were studied in combination with simulations from the 0-D Enthalpy Based Thermal Evolution of Loops (EBTEL) model.
} 
{
At the flare on-set, upflows of $\sim$90~km~s$^{-1}$ and low level emission were observed in \ion{Fe}{xix}, consistent with pre-flare heating and gentle chromospheric evaporation. During the impulsive phase, upflows of $\sim$80~km~s$^{-1}$ in \ion{Fe}{xix} and simultaneous downflows of  $\sim$20~km~s$^{-1}$ in \ion{He}{i} and \ion{O}{v} were observed, indicating explosive chromospheric evaporation. The plasma was subsequently found to reach a peak temperature of $\ga$13~MK in approximately 10 minutes. Using EBTEL, conduction was found to be the dominant loss mechanism during the initial $\sim$300~s of the decay phase. It was also found to be responsible for driving gentle chromospheric evaporation during this period. As the temperature fell below $\sim$8~MK, and for the next $\sim$4,000~s, radiative losses were determined to dominate over conductive losses. The radiative loss phase was accompanied by significant downflows of $\leq$40~km~s$^{-1}$ in \ion{O}{v}.
}
{
This is the first extensive study of the evolution of a canonical solar flare using both spectroscopic and broad-band instruments in conjunction with a hydrodynamic model. While our results are in broad agreement with the standard flare model, the simulations suggest that both conductive and non-thermal beam heating play important roles in heating the flare plasma during the impulsive phase of at least this event.
} 
{}

\authorrunning{Raftery et al.}
\maketitle
 
 \begin{figure*}[!hbt]

      \includegraphics[width=\textwidth, trim =-10 -40 -10 -30, clip = true]{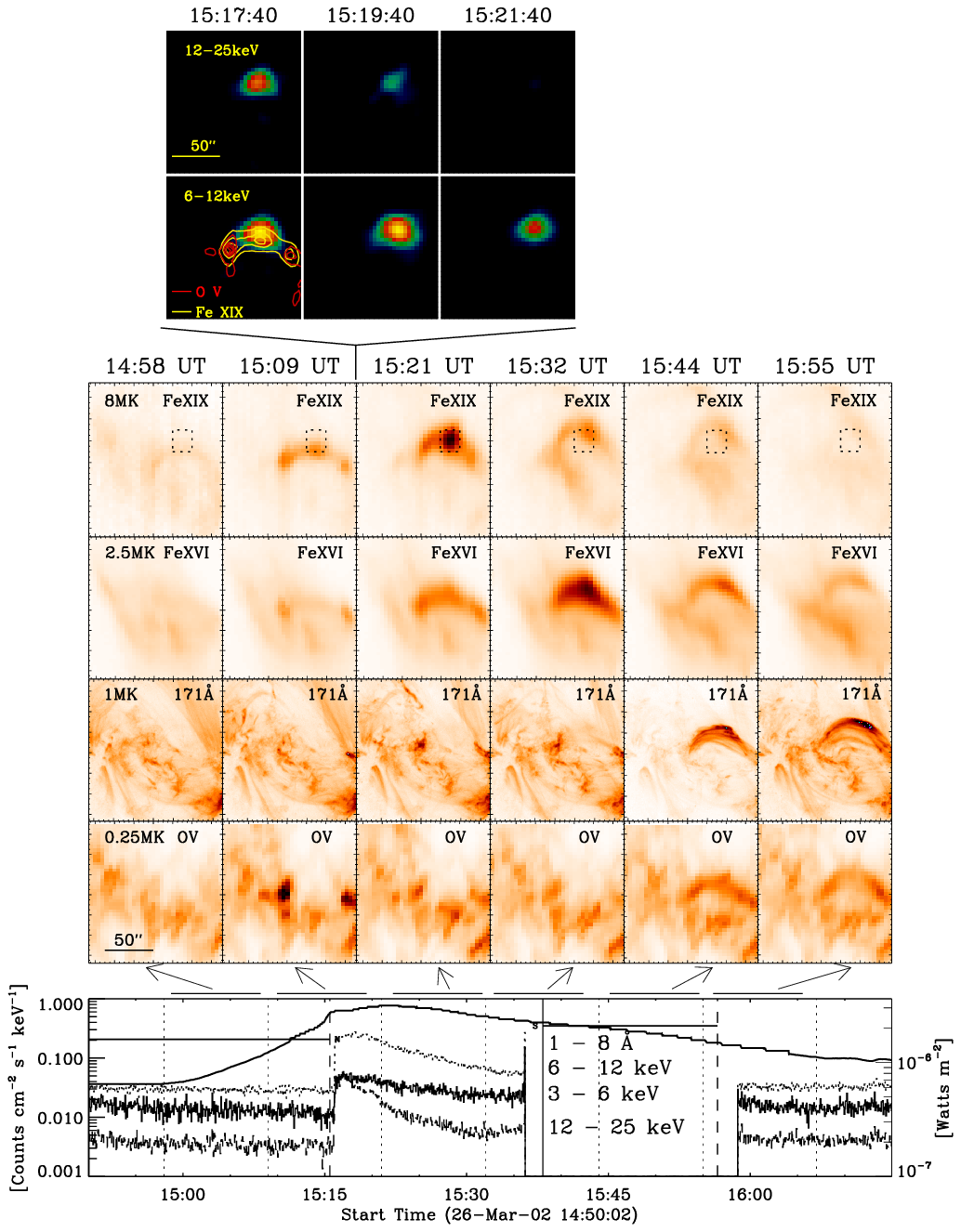}

\caption{Evolution of the GOES C3.0 flare observed on 2002 March 26 between 15:00~UT and 16:30~UT. RHESSI 6-12 and 12-25~keV images, integrated over two minutes (the time shown $\pm$1~minute) are shown in the top two panels. The contours from \ion{O}{v} and \ion{Fe}{xix} observed at 15:09~UT are overplotted in red and yellow respectively. \ion{Fe}{xix} (8MK), \ion{Fe}{xvi} (2.5~MK), TRACE/171~\AA \ (1.0~MK) and \ion{O}{v} (0.25~MK) emission is shown in the next four panels. The dotted box overplotted on the \ion{Fe}{xix} images represents the region of the loop used in the temperature and emission measure investigation. The bottom panel shoes the GOES 1--8~\AA, RHESSI 3--6, 6--12 and 12--25~keV lightcurves. RHESSI was in eclipse until 15:15~UT and passed through the South Atlantic Anomaly between 15:35~UT and 15:58~UT. The vertical dotted lines (and corresponding arrows) on the GOES plot represent the start and end times of the CDS rasters above. 
       }
   \label{fig:F-4panels}
\end{figure*}

\section{Introduction}

The temporal evolution of most solar flares can be divided into two distinct phases. During the impulsive phase temperatures rise to $\ga$10~MK via direct heating below the reconnection site in the corona and/or the process of chromospheric evaporation due to accelerated particles \citep{Kopp76}. Chromospheric evaporation driven by accelerated particles can be classified in one of two ways - explosive or gentle \citep{Fisher85, Milligan_explosive, Milligan_gentle}. Explosive evaporation occurs when the flux of non-thermal particles impacting the chromosphere is greater than a critical value ($\sim$3$\times10^{10}$~ergs~cm$^{-2}$~s$^{-1}$) and the chromosphere cannot dissipate the absorbed energy efficiently enough. The plasma is forced to expand into the corona as hot, upflows of hundreds of~km~s$^{-1}$ and simultaneously into the chromosphere as cooler downflows of tens of~km~s$^{-1}$. Beam driven gentle evaporation occurs when the non-thermal flux is less than $\sim$10$^{10}$~ergs~cm$^{-2}$~s$^{-1}$. Under these circumstances, the chromospheric response is efficient in radiating the absorbed energy. Gentle evaporation can also be driven by a downward heat flux from the corona. In both cases, plasma rises slowly (tens of~km~s$^{-1}$) upwards into the loop. Once the energy release has ceased, the hot plasma returns to its equilibrium state during the decay phase. The cooling process begins with thermal conduction as the dominant loss mechanism due to the high temperatures present. As the temperature decreases and the radiative loss function begins to increase, radiative cooling becomes more efficient \citep{Culhane1970}. Finally, the ``evaporated'' material drains back towards the solar surface, returning the system to equilibrium.

There have been a wealth of studies that focus on hydrodynamic modelling of these heating and cooling mechanisms (e.g. \citeauthor{Antiochos78},  \citeyear{Antiochos78}; \citeauthor{Fisher85}, \citeyear{Fisher85}; \citeauthor{Doschek83}, \citeyear{Doschek83}; \citeauthor{Cargill93}, \citeyear{Cargill93}; \citeauthor{Klimchuk01}, \citeyear{Klimchuk01}; \citeauthor{Reeves02}, \citeyear{Reeves02}; \citeauthor{Bradshaw05}, \citeyear{Bradshaw05}; \citeauthor{Klimchuk06}, \citeyear{Klimchuk06}; \citeauthor{Warren07}, \citeyear{Warren07}; \citeauthor{Sarkar08}, \citeyear{Sarkar08}). For example, \citet{Reale07} conducted an analysis of the details of stellar flares using the Palermo-Harvard theoretical model \citep{Peres82, Betta97}. This paper fully describes the cooling timescales and plasma parameters of flares in terms of their phases, including an investigation of the thermal heating function. However, these, and most other theoretical results were not compared to observations. The majority of investigations that make this comparison concentrate on broad-band instruments and utilise very simple models. For example \citet{Culhane1994} compared Yohkoh observations to an over-simplified power-law cooling curve. \citet{Aschwanden2001} compared broad-band observations to a model that considers a purely conductive cooling phase followed by a purely radiative cooling phase. \citet{Vrsnak06} conducted a similar study, again concentrating on a broad-band observations and a simple, independent cooling mechanism model. 

The work presented in this paper aims to improve on previous studies by comparing high resolution observations over a wide range of temperatures to a detailed theoretical model. Observations of a GOES C-class solar flare were made with several instruments, including the Transition Region and Coronal Explorer (TRACE, \citeauthor{Handy99} \citeyear{Handy99}), the Reuven Ramaty High Energy Solar Spectroscopic Imager (RHESSI; \citeauthor{Lin02} \citeyear{Lin02}), GOES-12 and the Coronal Diagnostic Spectrometer (CDS;  \citeauthor{Harrison95} \citeyear{Harrison95}). There are many advantages to using spectroscopic data in conjunction with broad-band observations. The identification of emission lines are, for the most part, well documented and therefore individual lines can be isolated for analysis. Also, material as cool as 30,000~K can be observed simultaneously with emission at 8~MK. Furthermore, it is possible to carry out velocity, temperature and emission measure diagnostics over a wide range of temperatures for the duration of the flare, significantly improving the scope of the analysis undertaken. These observations were compared to a highly efficient 0-D hydrodynamic model - the Enthalpy Based Thermal Evolution of Loops (EBTEL; \citeauthor{Klim07} \citeyear{Klim07}). 

The combination of this extensive data set and the new modelling techniques enables a comprehensive analysis of the heating and cooling of flare plasma to be performed. Section \ref{section:observations} describes the observations of this flare and the diagnostic tools used. Section \ref{section:modelling} contains a summery of the theoretical models and Sect. \ref{section:results} lays out the results of this study. The conclusions and future work are discussed in Sect. \ref{section:concs}.


\section{Observations and data analysis}
\label{section:observations}
This investigation concentrates on a GOES C3.0 flare that occurred in active region NOAA AR9878 on 2002 March 26 close to disk centre ($-$92\arcsec, 297\arcsec), beginning at $\sim$15:00~UT.
The CDS observing study used (FLARE\_AR) focused on five emission lines spanning a broad range of temperatures. The rest wavelengths and peak temperatures of each of the lines and those of the RHESSI, GOES and TRACE passbands are given in Table \ref{table:lam0}, where the quoted temperatures refer to the maximum of the response function.
Each raster consists of 45 slit positions, each 15~s long, resulting in an effective cadence of $\sim$11~minutes. The slit itself is 4$\arcsec \times$180$\arcsec$, resulting in a 180$\arcsec \times$180$\arcsec$ field of view.

Figure \ref{fig:F-4panels} shows the evolution of the flare in multiple wavelengths. The top two rows show the loop top source observed in RHESSI 6-12 and 12-25~keV energy bands, with the \ion{O}{v} and \ion{Fe}{xix} (15:09~UT) contours overplotted. The next four rows of this figure shows emission observed in \ion{Fe}{xix}, \ion{Fe}{xvi}, TRACE 171~\AA \ and \ion{O}{v}. The GOES 1--8~\AA \ and RHESSI 3--6, 6--12 and 12--25~keV lightcurves are shown at the bottom of the figure.


As Fig. \ref{fig:F-4panels} shows, at $\sim$15:00~UT, before the main impulsive phase of the flare began, evidence of low level \ion{Fe}{xix} loop emission was observed. By $\sim$15:09~UT, the footpoints were seen in \ion{O}{v} while the \ion{Fe}{xix} loop top emission continued to brighten. At 15:16:40~UT, when RHESSI emerged from eclipse, a thermal looptop source was observed in both 6-12 and 12-25~keV energy bands, with a corresponding temperature of $\sim$15~MK. By 15:21~UT the dominant emission had cooled to $\sim$8~MK. At this time, a bright ``knot'' can be seen at the top of the loop. Such features have been observed in the past and have not been readily explained (e.g. \citeauthor{Doschek05}, \citeyear{Doschek05}). During the early decay phase ($\sim$15:32~UT), the loop is seen to cool into the \ion{Fe}{xvi} temperature band ($\sim$2.5~MK) and by $\sim$15:44~UT, the plasma has cooled to $\le$1~MK, as seen by TRACE and in \ion{O}{v}. 

Although RHESSI was in eclipse for the majority of the impulsive phase of the flare, the observed continued rise of the 6--12~keV lightcurve after emergence from night, implies that the peak of the soft X-rays (SXR) was observed. However, while a hard X-ray (HXR) component was observed at this time, it is not believed to be the peak of non-thermal emission. 

\begin{table}[!h] 
\caption{Rest wavelengths and temperature of emission lines and bandpasses used in this study.} 

\centering 
\begin{tabular}{c c c} 
\hline\hline

Ion& $\lambda_{0}$~[\AA] & Temperature~[MK]\\
\hline
 \ion{He}{i} & 548.45 & 0.03 \\
 \ion{O}{v} & 629.80 & 0.25 \\
 \ion{Mg}{x} & 625.00 & 1.2 \\
 \ion{Fe}{xvi} & 360.89 & 2.5 \\
 \ion{Fe}{xix} & 592.30 &  8.0\\
 TRACE & 171 & 1.0 \\
 \hline
\hline
Instrument & Range & Temperature~[MK]\\
 \hline
 GOES & 0.5--4\AA \ \& 1--8\AA & 5 -- 30 \\
 RHESSI & 3~keV--17~MeV &  $\ga$5 \\
\hline
     \end{tabular} 

\label{table:lam0} 

\end{table}


\subsection{Temperature and emission measure}
\label{section:temperature}

The temperature and emission measure evolution for this flare was determined by analysing lightcurves from RHESSI, GOES, CDS and TRACE. 

\begin{figure} [!t]
\centerline{\hspace*{-0.05\textwidth}
 \includegraphics[width=0.5\textwidth, trim =0 100 35 0, clip = true]{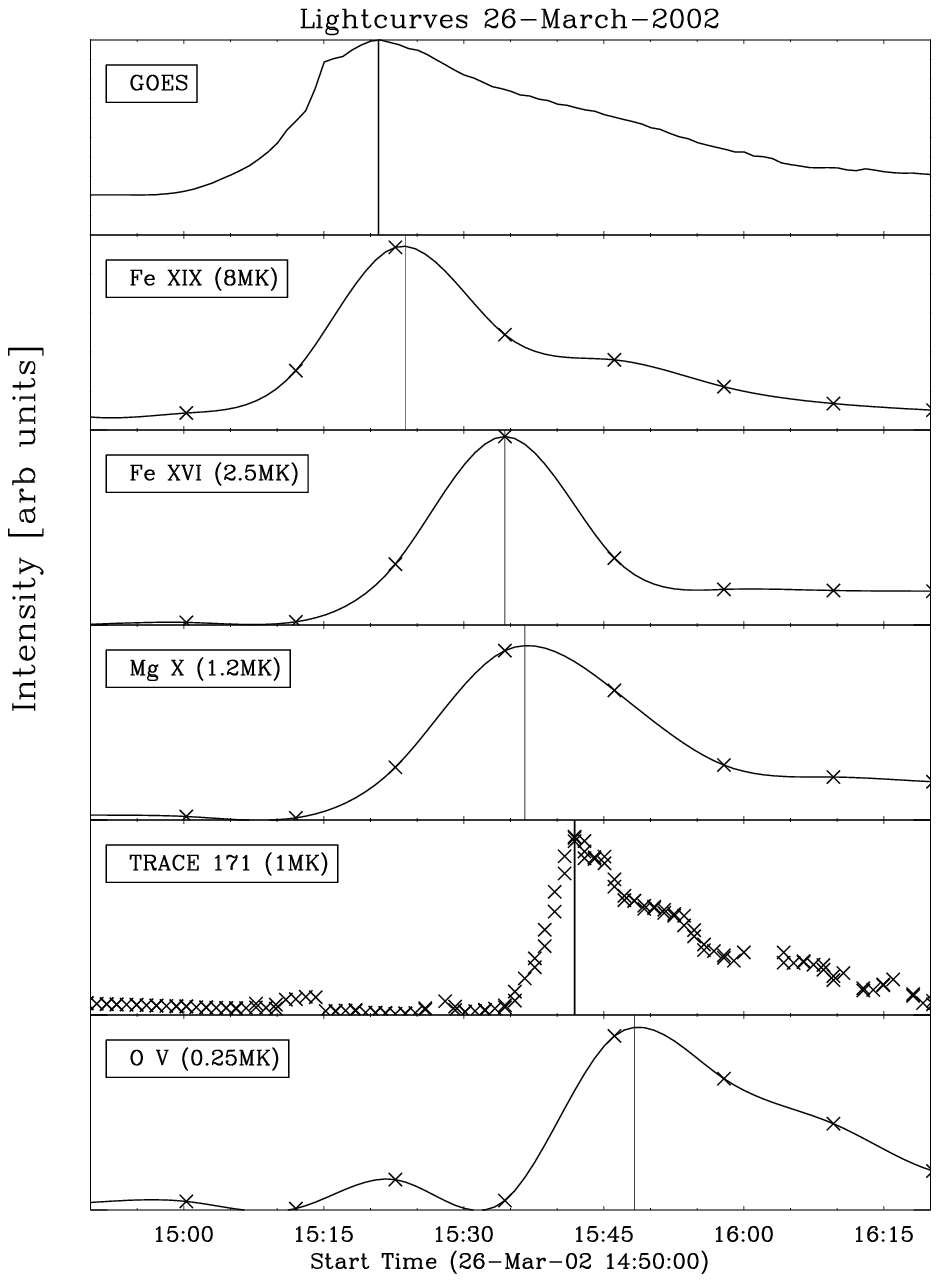} 
}
 \caption{\textbf{Linear} plot of the lightcurves of emission observed by GOES, CDS and TRACE. The data points were fit using a spline interpolation. The vertical represent the peaks of the lightcurve fits.   } 
 \label{fig:lgtcv} 
\end{figure}

\begin{figure} [!t]
\centerline{\hspace*{-0.05\textwidth}
 \includegraphics[width=0.5\textwidth, trim =0 170 0 0, clip = true]{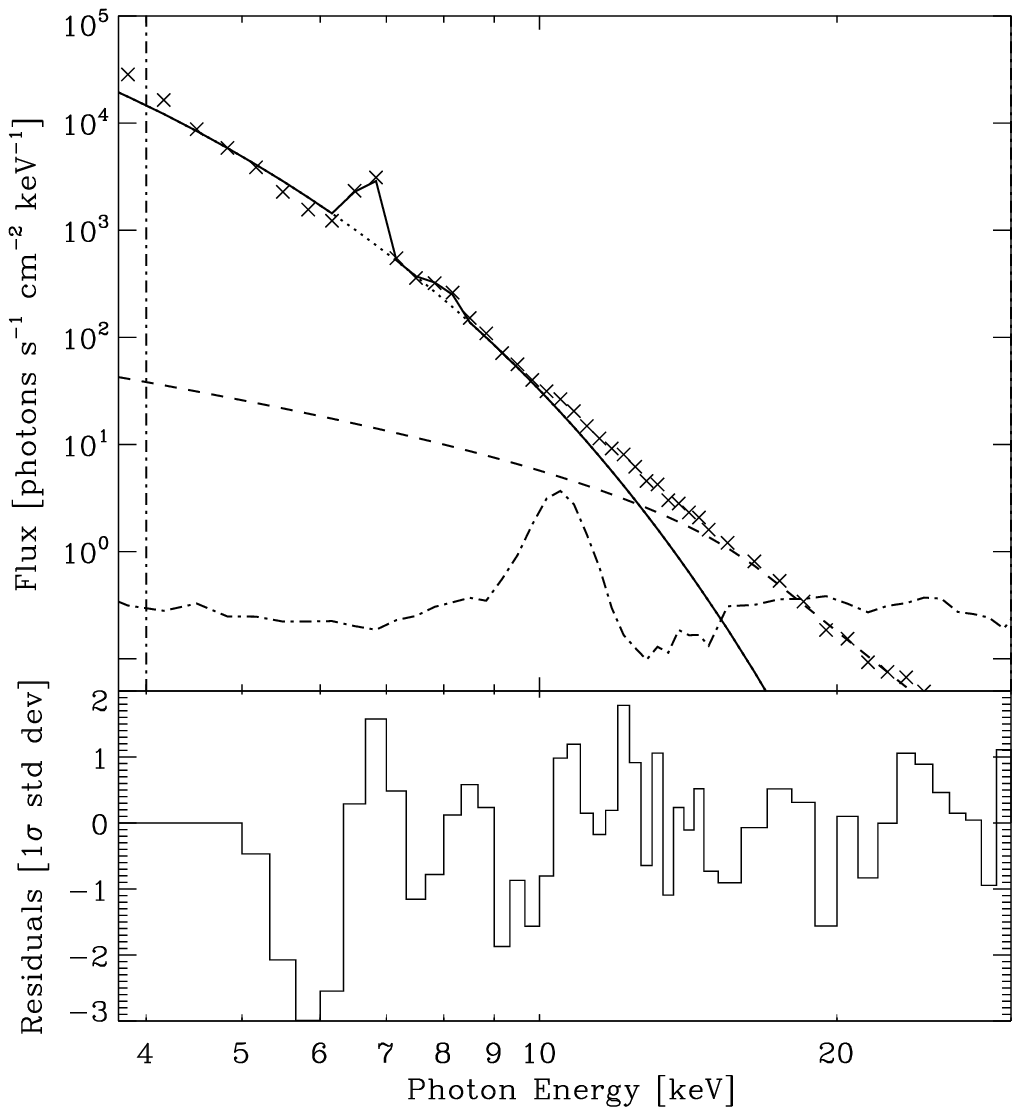} 
}
 \caption{The top panel shows the RHESSI photon spectrum between 15:16:30~UT and 15:17:30~UT. The residuals are shown in the bottom panel.   } 
 \label{fig:hsi} 
\end{figure}

The RHESSI spectrum, shown in Fig. \ref{fig:hsi}, was analysed over one minute at the peak of the 6--12~keV energy band. Following previous studies (e.g. \citeauthor{Saint-Hiliare02}, \citeyear{Saint-Hiliare02}), the data were fitted with an isothermal model at low energies and a thick-target model up to $\sim$30~keV. The thick target component yielded a low energy cutoff of 17~keV and a power law index of 8.2. The isothermal fit to lower energies determined a temperature and emission measure of $\sim$13~MK and  $\sim$1$\times$10$^{48}$~cm$^{-3}$ respectively. A non-thermal electron flux of $\sim$7$\times$10$^{9}$~ergs~cm$^{-2}$~s$^{-1}$ was also calculated, approximating the footpoint area from \ion{He}{i} and \ion{O}{v} observations. Since it is probable that the HXR peak occurred before this time, this value is taken to be a lower limit to the maximum non-thermal electron flux. Following \citet{White05}, the filter ratio of the two GOES passbands produced a temperature profile, giving a peak temperature of 10~MK and an emission measure of 4$\times$10$^{48}$~cm$^{-3}$.

CDS and TRACE~171~\AA \ observations were integrated over a fixed area described by the bright ``knot'', seen in \ion{Fe}{xix} at 15:21~UT in Fig. \ref{fig:F-4panels}. This area is highlighted by the dotted box in the \ion{Fe}{xix} images. The lightcurves of this region were analysed for each raster and TRACE image during the course of the flare (Fig. \ref{fig:lgtcv}). The loop is believed to consist of multiple magnetic strands. It is assumed that the majority of strands within the region of this ``knot'' are heated almost simultaneously at the time of the HXR burst. A small number of strands can be heated before or after this time, producing a multi-thermal plasma. However, this small region is approximately isothermal at any one time. For the remainder of this paper, any reference to the ``loop apex'' refers to the area defined by the dotted box on the \ion{Fe}{xix} images in Fig. \ref{fig:F-4panels}. The \ion{He}{i} line was not used for the determination of temperature as it is optically thick and neither \ion{He}{i} nor TRACE were used in the emission measure analysis.

\begin{figure} [!t]
\centerline{\hspace*{-0.02\textwidth}
 \includegraphics[width=0.5\textwidth, trim =0 80 25 40, clip = true]{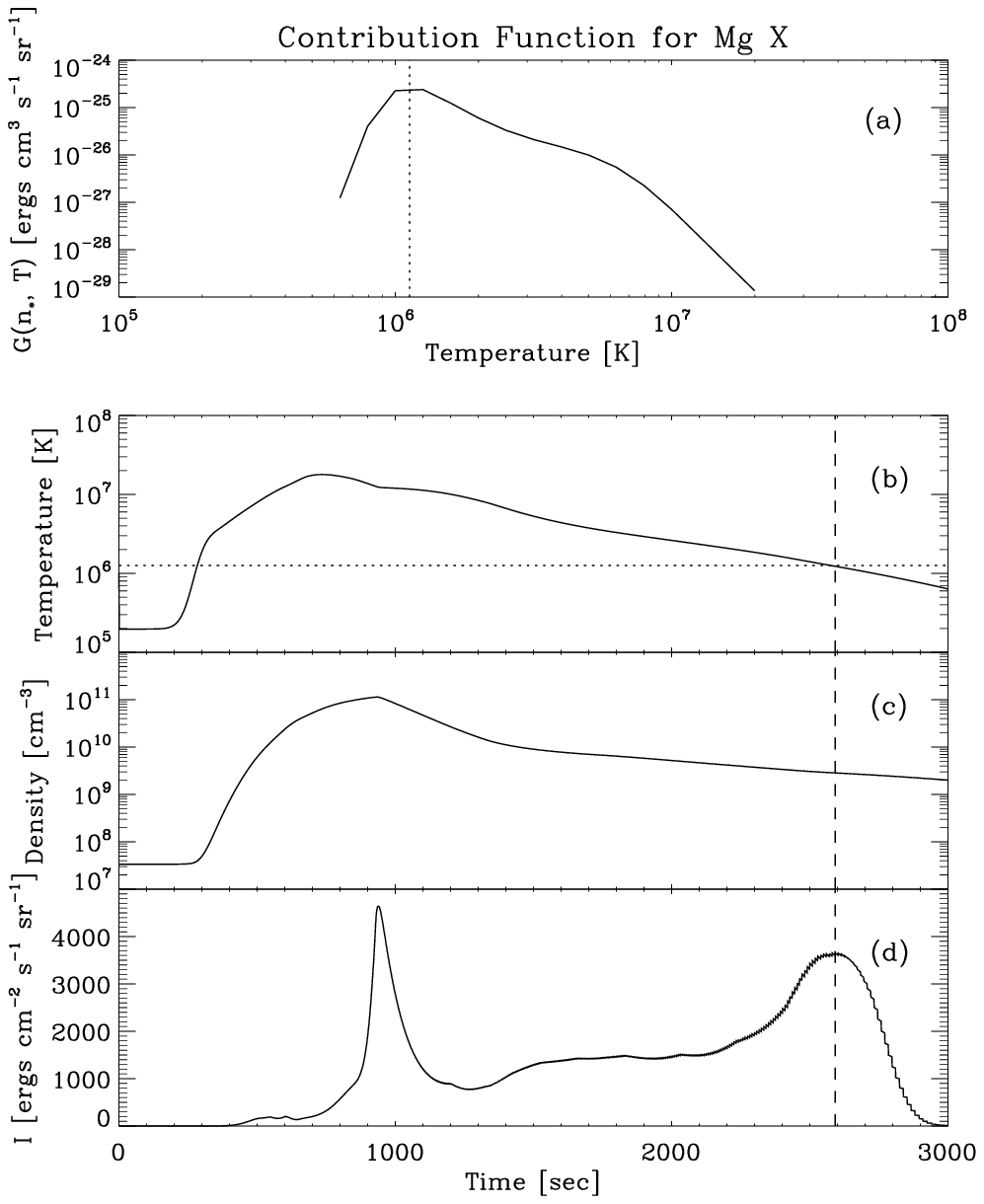} 
}
 \caption{Panel (a) shows the \ion{Mg}{x} contribution function calculated from CHIANTI. The peak is highlighted with a dotted vertical line. This occurs at 1.25~MK. Panel (b) shows the simulated temperature profile calculated by the EBTEL model (see Sect. \ref{section:modelling} for details). The dotted horizontal line denotes the temperature of the contribution function peak from panel (a), and the dashed vertical line is the time this temperature is reached. Panel (c) shows the simulated density profile, and panel (d) shows the simulated lightcurve for \ion{Mg}{x}, calculated from Eqn. \ref{equation:lgtcv}. 
  \label{fig:peak_proof} 
  }
  \vspace*{-0.01\textwidth}

\end{figure}

To obtain the thermal evolution of this flare, the time of the peak of each lightcurve was assigned the associated temperature mentioned in Table \ref{table:lam0}. The uncertainty in the temperature measurement was taken to be the width of the appropriate contribution function. 
The motivation for this assumption is indicated in Fig. \ref{fig:peak_proof}. Figure \ref{fig:peak_proof}a shows the contribution function for \ion{Mg}{x} calculated using CHIANTI version 5.2 \citep{Dere97, Landi06}. Figure \ref{fig:peak_proof}b shows a theoretical temperature evolution for a flare with a maximum of $\sim$20~MK and Fig. \ref{fig:peak_proof}c shows the corresponding density evolution. The lightcurve (Fig. \ref{fig:peak_proof}d) was calculated by taking the density and temperature at each time step, and calculating the value of the contribution function for that particular temperature and density. The intensity, $I$, was calculated using:
\begin{equation}
\label{equation:lgtcv}
I =\int G{(n_{*}, T)} n_{e}^{2}\,dV = G{(n_{*}, T)} EM = G{(n_{*}, T)} n_{e}^{2} f V
\end{equation}
for constant volume, $V$, temperature $T$, filling factor, $f$, electron density, $n_{e}$ and emission measure $EM$. $G{(n_{*}, T)}$ is the contribution function calculated for constant density, $n_{*}$\footnote{A study of the contribution function dependence on density was carried out. It was found that varying $n_{*}$ by two orders of magnitude resulted in little or no change to the contribution function. This density was assigned a constant of value of $10^{12}$~cm$^{-3}$.} and variable temperature, $T$. During the impulsive phase, as the corona is heated, the temperature jumps by $\sim$2 orders of magnitude, passing quickly through the peaks of the contribution function and density curves. As this happens, a sharp peak is seen in the lightcurve at $\sim$1000~s. As the plasma cools, a broader peak is seen in the lightcurve, relating to emission from the looptop. This occurs at $\sim$2600~seconds (dashed lines, Fig. \ref{fig:peak_proof}b, c, d), a time corresponding to a temperature of 1.25~MK, the temperature at which the contribution function is maximised (dotted lines, Fig. \ref{fig:peak_proof}a, b). Therefore, the lightcurve peak seen in the cooling phase of the flare can be attributed to passing through the maximum of the contribution function; i.e., at 2600~s, the plasma is emitting at 1.25~MK. 

Under the assumption of an isothermal plasma, the emission measure was obtained at the times of the lightcurve peaks for \ion{O}{v}, \ion{Mg}{x}, \ion{Fe}{xvi} and \ion{Fe}{xix}, following Eqn. \ref{equation:lgtcv}. We have not included TRACE in the analysis of EM due to the ill-defined instrument response of the TRACE 171 \AA \ band. The uncertainty in calculating EM using RHESSI was found to be approximately 50\% using a thermal fit to the spectrum. This is in agreement with the values found by the RHESSI instrument team \citep{McTiernan06}. 

For CDS, there are a number of factors to consider. These in include uncertainties in the intensity of the line, the contribution function and the CDS calibration. While the uncertainty in measuring the line intensity is small for strong lines such as those used in our study, typically $\sim$10\% \citep{DelZanna01}, a consideration of the contribution functionÕs FWHM results in an uncertainly in the EM of approximately 30\%. The CDS calibration is also known to be good to within 15-20\% \citep{Brekke00}. The GOES instruments is known to have limited ability for making accurated measurements of both temperature and EM. Accepted values from \citet{Garcia00}, give the EM uncertainty to be 10\%. Considering these factors, the combined photometric error for all instruments was taken to be 50\%.

\subsection{Velocity}
\label{section:velocity}
\begin{figure}[!t]
               \includegraphics[width=0.4\textwidth,trim =20 350 5 0, clip = true]{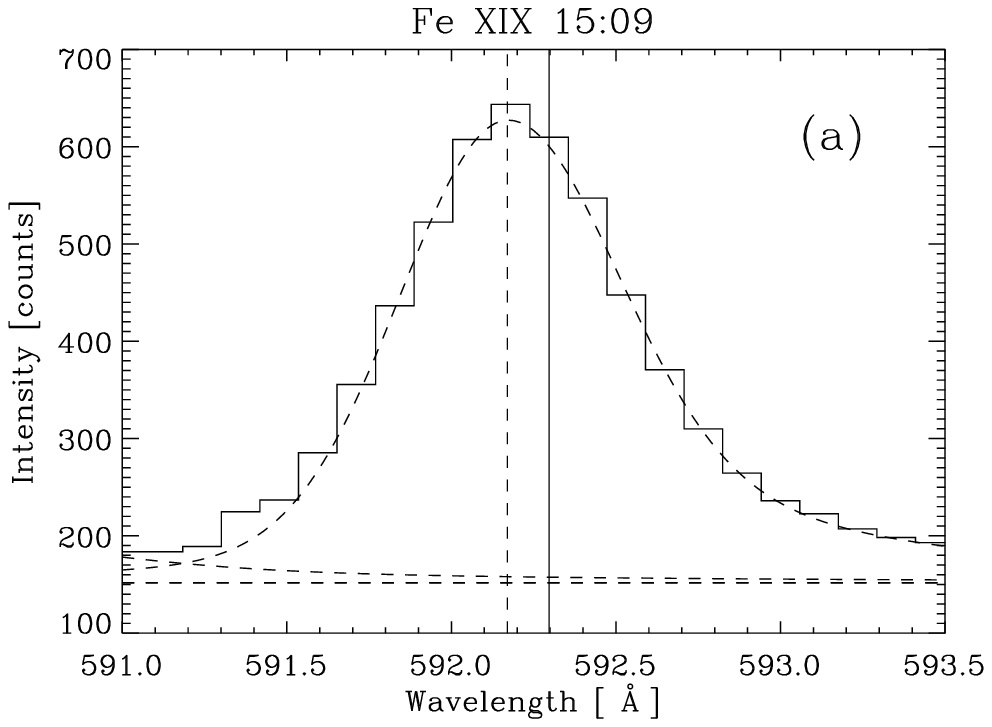}
               \includegraphics[width=0.4\textwidth,trim =20 350 5 0, clip = true]{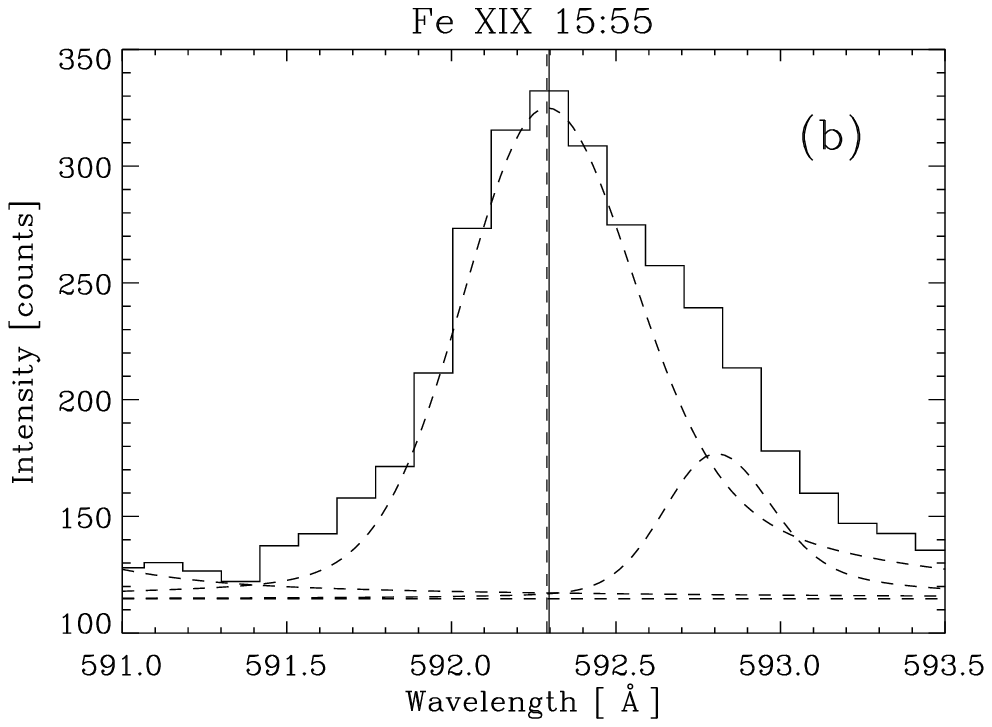}

\caption{The line profiles of the \ion{Fe}{xix} emission line for the right footpoint during the impulsive phase (a) and the decay phase (b). Note the increasing intensity of the \ion{Fe}{xii} blend at $\sim$592.8~\AA \ \citep{DelZanna05} during the decay phase. The solid vertical line represents the rest wavelength while the dashed vertical line is the centroid of the line.         }

   \label{fig:line_prof}
   \end{figure}

The Doppler shifts at both footpoints were calculated for the duration of the flare using the five CDS emission lines, with uncertainties of $\pm10$~km~s$^{-1}$ \citep{Brekke97, Gallagher99}. The rest wavelengths for \ion{He}{i}, \ion{O}{v}, \ion{Mg}{x} and \ion{Fe}{xvi} were calculated from a region of pre-flare quiet Sun. For the \ion{Fe}{xix} line, a detailed analysis of the behaviour of wavelength as a function of time was conducted to establish the rest wavelength. Centroid wavelengths from both before the SXR rise and from late in the flare were averaged and corrected for both heliocentric angle and an average inclination of 44$^{\circ}$ to obtain the rest wavelength. Fig. \ref{fig:line_prof} shows the \ion{Fe}{xix} line profiles for the right footpoint during the impulsive (a) and decay (b) phases.

\section{Modelling}
\label{section:modelling}

\subsection{The Cargill model}
\label{section:cargill}

Following \citet{Antiochos76}, \citet{Cargill93, Cargill94} presented a model that considered a flare that is cooling purely by conduction for a time $\tau_{c}$, followed by purely radiative cooling for a time $\tau_{r}$. The cooling time-scales were given by:

\begin{equation}
\label{equation:con_cool_time}
\tau_{c} = 4 \times 10^{-10}\frac{nL^{2}}{T^{5/2}}, 
\end{equation}
and
\begin{equation}
\label{equation:rad_cool_time}
    \tau_{r} = \frac{3kT}{n \Lambda{(T)}}
\end{equation}
where $\Lambda{(T)}$ is the radiative cooling function (see  \citeauthor{RTV78} \citeyear{RTV78} for further details), $L$ is the loop half length, and all other variables have their usual meaning.

 The time and temperature at which the cooling mechanism dominance changes, $\tau_{*}$ and $T_{*}$ respectively, can be calculated as follows: 

\begin{equation}
\label{equation:crit_time}
\tau_{*} = \tau_{c_{0}} \left[ \left( \frac{\tau_{r_{0}}}      {\tau_{c_{0}}}   \right)^{7/12} -  1 \right],
\end{equation}
and
\begin{equation}
\label{equation:crit_time}     
T_{*} = T_{i}\left( \frac{\tau_{r} }{\tau_{c}}\right)^{-1/6}
\end{equation}

where $T_{*} = T(t = \tau_{*})$ and subscript $0$ denotes initial values. 
These parameters were calculated for the flare under investigation and compared to the results of the 0-D hydro model.

\subsection{The EBTEL model}

Enthalpy Based Thermal Evolution of Loops (EBTEL) model is a 0D model that simulates the evolution of the average temperature, density, and pressure along a single strand \citep{Klim07}, calculating a single value of each of these quantities at any given time. This is a reasonable representation since temperature, density, and pressure are approximately uniform along the magnetic field, with the exception of the steep gradients in the transition region at the base of the loop.
 
As its name implies, EBTEL takes explicit account of the important role of enthalpy in the energetics of evolving loops. Under static equilibrium conditions, less than half of the energy deposited in the corona is radiated directly. The rest is thermally conducted down to the transition region, where it is radiated away. Under evolving conditions, chromospheric evaporation occurs when the transition region cannot accommodate the downward flux, or, if the flux is insufficient to power the transition region radiative losses, condensation occurs. 

  EBTEL equates an enthalpy flux with the excess or deficit heat flux. Kinetic energy is ignored because the flows are generally subsonic, except perhaps in the earliest times of an impulsive event. Another assumption made is that the radiative losses from the transition region and corona maintain a fixed proportion at all times. EBTEL has been compared with sophisticated 1D hydrodynamic models and found to give similar results, despite using 4 orders of magnitude less computing time.

EBTEL allows for any temporal profiles of both direct plasma heating and non-thermal particle acceleration. The effects of the non-thermal electron beam are treated in a highly simplified manner.  It is assumed that all of the energy goes into evaporating plasma. This is reasonable for gentle evaporation \citep{Fisher85}, but for explosive evaporation, some of the beam energy will go into a plug of downflowing and radiating plasma deep in the chromosphere. Thus the actual energy of the beam, as inferred from RHESSI observations for example, is greater than the beam energy used in the EBTEL simulation. For a complete description of this model, refer to \citet{Klim07}.
	
The flare loop is almost certainly composed of many strands that are heated at different times. However, the observations suggest that most of the strands are heated in approximately the same way and at approximately the same time (i.e. during the HXR burst), so the flare was modelled as a single monolithic loop. Nonetheless, some strands are expected to be heated both before and after this main bundle (e.g., \citeauthor{Klimchuk_soho06}, \citeyear{Klimchuk_soho06}), and this will result in some deviations between the model and observations.

\begin{figure*}[!ht]
\centerline{\hspace*{0.0\textwidth}
 \includegraphics[width=0.8\textwidth, trim =0 0 0 0, clip = true]{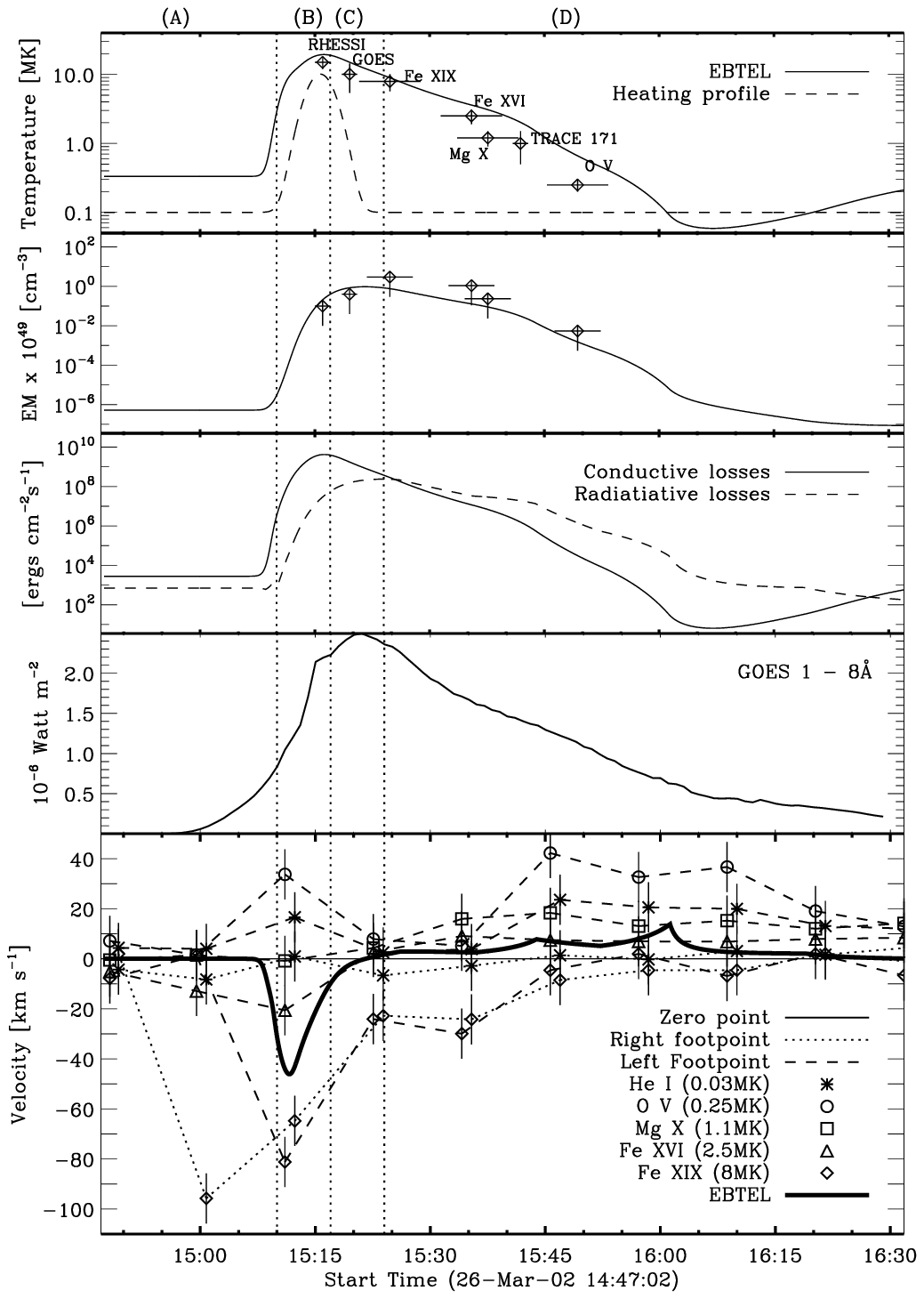} 
}
 \caption{The EBTEL temperature evolution (solid line) that best reproduced the observations of Sect. \ref{section:temperature} is shown in the first panel, along with the heating function whose parameters are described in Table \ref{table:params} (dashed line, arbitrarily scaled to show position and width). The second panel shows the corresponding model and observed emission measure evolution, with the data points corresponding in time to those in the first panel. The third panel shows the conductive and radiative losses throughout the flare on solid and dashed lines respectively. The fourth panel shows the GOES 1 -- 8~\AA \ lightcurve. The bottom panel shows the velocities of the right footpoint for the five CDS lines and that of the left footpoint in \ion{He}{i} and \ion{Fe}{xix}. The velocities were very similar at both footpoint in all other lines and so were omitted for clarity. The EBTEL simulated velocity for \ion{Mg}{x} is represented by the thick black line. The dotted vertical lines correspond to the flare phases (A)--(D) explained in Sect. \ref{section:phases}.} 
 \label{fig:full} 
 \vspace{-0.1cm}
\end{figure*}

	\begin{table*}
\caption{Input parameters used for modelling simulation. The parameters were constrained by data when possible and the ranges of parameters investigated are shown. } 

\centering 
\begin{tabular}{ l  c  c  c  c  } 
\hline\hline
	
Parameter 						 								& Observed 					&		EBTEL					\\
\hline
 Loop half-length [cm] 												& 3$\times$10$^{9}$			&		$(3 \pm 0.2)\times10^{9}$	\\
 Non-thermal flux  													&							& 								\\
 							- Amplitude [ergs~cm$^{-2}$~s$^{-1}$]		& 7$\times$10$^{9}$			&		$5 \times 10^{8\pm1}$				\\
							- Width [sec]							& $\sim$100					&		$100\pm50$				\\
							- Total  [ergs~cm$^{-2}$]					& $\sim$1.7$\times$10$^{12}$		&		$2.5\times10^{10\pm1}$		\\
 Direct  heating rate												 	& 							&								\\	
 							- Amplitude [ergs~cm$^{-3}$~s$^{-1}$]		& -							&		$0.7\pm 0.3$				\\
							- Width [sec]							& -							&		$100\pm 50$				\\
							- Background [ergs~cm$^{-3}$~s$^{-1}$]		& -							&		$\le1\times10^{-6}$		 	\\
							- Total  [ergs~cm$^{-3}$]					& -							&		$175\pm150$				\\					 
Direct/non-thermal heating											&	\textbf{	(best fit parameters)		}			&		$\sim4$				\\

 \hline
     \end{tabular} 

\label{table:params} 

\end{table*}

The pre-flare conditions included a temperature of 0.3~MK, an initial density of 5$\times$10$^{7}$~cm$^{-3}$ and an emission measure of 4$\times$10$^{43}$~cm$^{-3}$. Input values were, where possible, constrained by observations. The loop length was determined from magnetic field extrapolations of the region (Conlon, \emph{priv. comm.}). The filling factor could not be constrained by data due to the poor spatial resolution of the high temperature instruments. Since it was assumed that the heating function is associated with the HXR burst, the majority of which was not observed, the shape of the heating function was inferred from previous observations of HXR bursts and the slow rise of the GOES SXR lightcurve. The most appropriate heating function was Gaussian in shape. The non-thermal electron flux was constrained by that found by the lower limit calculated from RHESSI observations and the width was inferred from the derivative of the SXR flux \citep{Neupert68, Zarro93}. While the direct heating rate was not constrained by observations, it was assumed to have the same width as the non-thermal heating flux and to occur at the same time. Due to the sensitivity of the model parameters to cooling timescales, the cooler data points (e.g. \ion{Fe}{xvi}, \ion{Mg}{x}, TRACE and \ion{O}{v}) were critical in constraining the parameters. The ranges of acceptable parameter values are shown in Table \ref{table:params}. The values obtained from observations are shown, along with the parameter values used in producing the results in Sect. \ref{section:results}. The range of parameter values shown correspond to the maximum and minimum values that produce an acceptable fit to data. The ratio of the heating components (i.e. direct to non-thermal) is also shown for the best fit parameters presented in Sect. \ref{section:results}, where the equivalent direct energy flux is given by the volumetric heating rate divided by the half loop length.

\section{Results}
\label{section:results}

Combining the observations from the different instruments used for this study, and the results from EBTEL, the heating and cooling phases of this flare can be comprehensively described. These results are presented in Figs. \ref{fig:full} and \ref{fig:flare_life}. These figures show  the evolution of the flare through the dependence of temperature, emission measure, energy losses and velocity, as discussed in Sects. \ref{section:temperature} and \ref{section:velocity}.

\subsection{Comparison of model to data}

The top two panels of Fig. \ref{fig:full} describe the evolution of the flare temperature and emission measure, respectively. The data points for each instrument were obtained using the analysis described in Sect. \ref{section:temperature}. The input parameters for EBTEL were approximated by observations and allowed to vary slightly until a good correlation with the cooling phase data was obtained. The conductive and radiative loss curves generated by EBTEL for the flare are shown in the third panel of Fig. \ref{fig:full}. Consistent with previous observations, conduction was found to dominate initially, with radiation becoming prevalent for the remainder of the decay phase. Both Cargill and EBTEL found conduction to dominate for the first 200 - 400 s of the decay phase, with radiation dominating for the remaining $\sim$4000 s, referring to $\tau_{c}$ and $\tau_{r}$ respectively. The time $\tau_{*}$ at which $\tau_{c} \approx \tau_{r}$ (i.e. the dominant loss mechanism switches from conduction to radiation) is $\sim$15:24~UT in both cases. However, the temperature, $T_{*} = T(\tau{*})$, at which this occurs was found to be $\sim$12~MK and $\sim$8~MK according to Cargill and EBTEL respectively. This discrepancy is due to the different approach to the modelling of the early decay phase. EBTEL simultaneously calculates the conductive and radiative losses throughout a flare while Cargill assumes cooling exclusively by either conduction or radiation at any one time. The fourth panel of this figure shows the GOES 1--8~\AA \ lightcurve for context. The last panel shows the velocities at the loop footpoints, calculated following the analysis in Sect. \ref{section:velocity}. The flow velocity at both left and right footpoints are shown for the coolest and hottest lines -- \ion{He}{i} and \ion{Fe}{xix} respectively, while for clarity, only the right footpoints for the remaining three lines were shown. The MgX Doppler shift simulated by EBTEL is represented by the thick black line. The simulations are in reasonable agreement with the observations. Upflows are of course predicted during the evaporation phase and downflows are predicted during the draining phase. However, the magnitudes are generally larger than those observed for reasons that we do not fully understand. Uncertainties in the velocity zero point adopted for the observations may account for the downflow discrepancy.  

The redshifts observed in the cooler lines during phase B, if real, are likely an indication of a downflowing chromospheric plug that accompanies explosive evaporation. The blueshifts seen in the hotter lines around 15:30~UT suggest that some loop strands were impulsively heated after the primary flare energy release. These blueshifts are expected to be smaller than the actual upflows because the spatially unresolved line profile represents a mixture of upflowing and downflowing strands \citep{Patsourakos06}. The small peaks seen at later times in the simulation velocity curve are a result of the piecewise continuous form used for the radiative loss function.

%

%

%

\begin{figure} 
\centerline{\hspace*{0.0\textwidth}
 \includegraphics[width=0.5\textwidth, trim =5 350 0 0, clip = true]{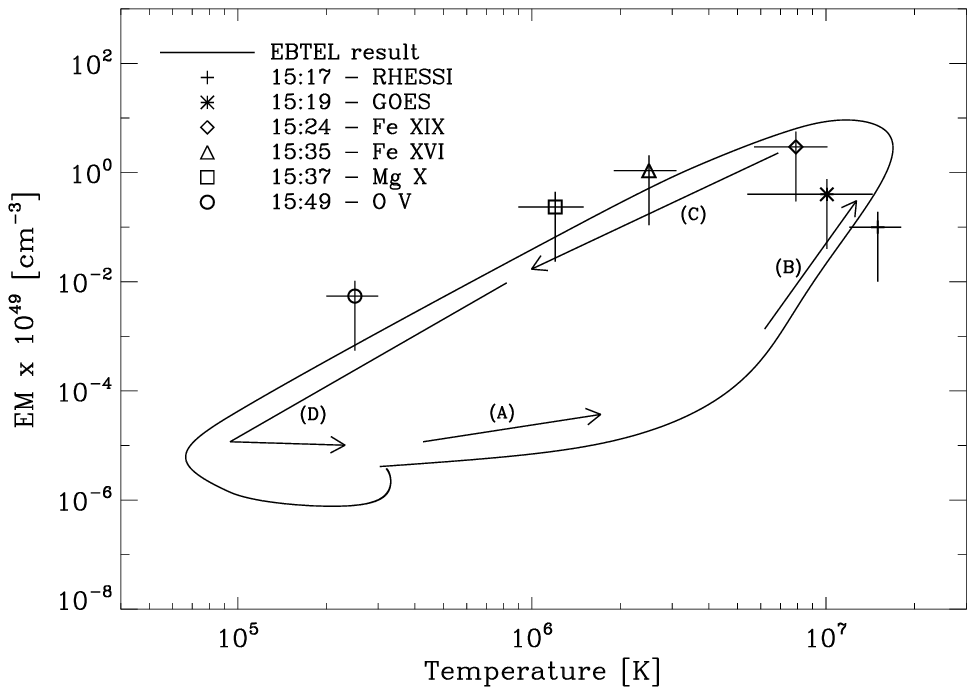} 
}
 \caption{This shows the dependence of emission measure on temperature for both model and data. The different phases of a solar flare are marked (A)--(D). Over-plotted are the emission measure data-points calculated in Sect. \ref{section:temperature} as a function of their temperature.  } 
 \label{fig:flare_life} 
 \vspace{-0.1cm}
\end{figure}

Figure \ref{fig:flare_life} shows the evolution of the flare through the interdependence of emission measure on temperature. The data points obtained during the analysis described in Sect. \ref{section:temperature} were computed at the same instant in time for any one emission line or bandpass. This figure shows the heating of the plasma (A) followed by evaporation of hot plasma (B), cooling (C) and draining (D). The plasma initially cooled to a temperature below the pre-flare value and asymptotically returned to the equilibrium state due to the low level, constant background heating.

\subsection{Flare Phases}
\label{section:phases}
Figures \ref{fig:full} and \ref{fig:flare_life} have sections labelled (A)--(D) which refer to the different phases of the flare, from pre-flare heating to the late decay phase and are described in detail in this section.

\begin{inparaenum}
\item[\itshape A\upshape)]
\emph{14:45--15:10; Pre-flare phase:} For the majority of this phase, the EBTEL parameters remained at quiet Sun values, as  Fig. \ref{fig:full} shows. At 15:07~UT the EBTEL temperature and emission measure began to rise. Figure \ref{fig:flare_life} shows the steep temperature gradient and the initial gradual rise in emission measure. However, as the fourth panel in Fig. \ref{fig:full} shows, the GOES soft X-rays began to rise slowly before this. At $\sim$15:00~UT, a small amount of \ion{Fe}{xix} emission was seen in the loop (Fig. \ref{fig:F-4panels}) and velocities of $\sim$90~km~s$^{-1}$ observed in \ion{Fe}{xix} can be seen in the bottom panel of Fig. \ref{fig:full} while all of the cooler lines remain at rest. This is evidence that pre-flare heating is driving gentle chromospheric evaporation in a small number of strands heated before the HXR burst.

\item[\itshape B\upshape)]
\emph{15:10--15:17; Impulsive phase:} During the impulsive phase of a flare, the standard model predicts the propagation of non-thermal electrons to the chromosphere where they heat the ambient plasma, causing it to rise and fill the loop \citep{Kopp76}. Upflows of $\sim$80~km~s$^{-1}$ in \ion{Fe}{xix} and simultaneous cool, downflows of $\sim$20~km~s$^{-1}$ in \ion{He}{i} and \ion{O}{v} were observed and shown in the bottom panel of Fig. \ref{fig:full}. A non-thermal electron flux of $\sim$7$\times10^{9}$~ergs~cm$^{-2}$~s$^{-1}$ was determined between 15:16:30 and 15:17:30~UT. This is slightly lower than the 3$\times10^{10}$~ergs~cm$^{-2}$~s$^{-1}$ required to drive explosive chromospheric evaporation \citep{Fisher85, Milligan_explosive}. As such, this value is taken to be a lower limit and that the maximum value of non-thermal flux occurred before 15:16~UT. 

\item[\itshape C\upshape)]
\emph{15:17--15:24; Soft X-ray peak:} The top panel of Fig. \ref{fig:full} shows the temperature has peaked and begun to fall and that the emission measure and SXRs were at a maximum in this phase. As the third panel of Fig. \ref{fig:full} shows, during this phase, conduction was efficiently removing heat from the corona and transferring it to the chromosphere, driving slow upflows of hot emission \citep{Zarro88}. These flows $\sim$20~km~s$^{-1}$ in \ion{Fe}{xix} respectively can be seen in the bottom panel of Fig. \ref{fig:full}.

\item[\itshape D\upshape)]
\emph{15:25--16:30; Decay phase:} This phase is dominated by radiative cooling, as seen in the third panel of  Fig. \ref{fig:full}. Velocities in \ion{Fe}{xix} were returning to quiet Sun values. The modest blueshift observed at 15:32~UT in \ion{Fe}{xix} suggests that the line profile contains components from evaporating strands that were heated after the main loop bundle. Between approximately 15:45 and 16:20~UT \ion{Mg}{x}, \ion{O}{v} and \ion{He}{i} showed downflows of up to $\sim$40~km~s$^{-1}$. This implies loop draining was occurring \citep{Brosius03}. By the end of the simulation, all of the parameters had returned to quiet Sun values. 

\end{inparaenum}

\section{Conclusions and Discussion}
\label{section:concs}
This paper compares a flare observed with CDS, TRACE, GOES and RHESSI to the 0-D hydrodynamic model EBTEL. Early in the impulsive phase of the flare, evidence of 8~MK emission and $\sim$90~km~s$^{-1}$ upflows suggest pre-flare gentle chromospheric evaporation. During the impulsive phase, hot upflowing plasma at velocities of $\sim$80~km~s$^{-1}$ and cool downflows of $\sim$20~km~s$^{-1}$ imply explosive chromospheric evaporation. Around the time of the soft X-ray peak, conduction was found to be highly efficient. Upflowing plasma at velocities of $\sim$20~km~s$^{-1}$ are observed in \ion{Fe}{xix}, suggesting conduction driven gentle chromospheric evaporation. The cooling timescales modelled by \citet{Cargill93, Cargill94} were tested against EBTEL and proved to agree reasonably well. The dominant cooling mechanism was found to switch from conduction to radiation at $\sim$15:24~UT in both models. However, the temperature at which this occurs ($\sim$8~MK and $\sim$12~MK for EBTEL and Cargill respectively) does not agree. This is as a result of the simultaneous cooling by conduction and radiation for EBTEL versus the exclusive cooling phases predicted by \citet{Cargill93, Cargill94}. The late decay phase of the flare is dominated by radiative cooling. Downflowing plasma observed in \ion{He}{i}, \ion{O}{v} and \ion{Mg}{x} during the late decay phase provides evidence of loop draining. 

By tracking the behaviour of this flare as it cooled through the response functions of the many instruments and emission lines, the evolution of the temperature and emission measure could be assessed. This evolution was then recreated using the EBTEL model, providing precise details, such as the cooling timescales and mechanisms, that cannot be easily obtained from data. For this particular flare, since the HXR burst was not fully observed, the details of the heating function could be also estimated from simulations. The description of the flare using both data and model allows for a much greater understanding of flare dynamics. The nature of these explosive events remain somewhat ambiguous, however further studies of this nature, will help to improve the understanding of them. 

The ratio of the heating functions were investigated. It was found that the observations were best modelled when the plasma was heated approximately equally by direct and non-thermal mechanisms. This implies that both of these processes are vital during the flaring process and that flares may not be energised primarily by non-thermal particles, as previously believed \citep{Brown71}. This is in agreement with recent results found by \citet{Milligan08}. There it was shown that a non-thermal electron beam is not necessarily required to obtain the high-temperature, high-density material we see in flares. However, it should be noted that the EBTEL value of the flux of non-thermal electrons required for equal heating is below the critical value for explosive evaporation hypothesised by \citet{Fisher85}. This can be explained by the over-simplifed treatment of non-thermal particles by EBTEL. This requires caution for a flare of this nature, where it is evident that non-thermal particles play an important role. The model assumes that all non-thermal energy is used for evaporating plasma upward into the loop. However, this may not be entirely true. It is well known that a very small fraction of this energy is used to produce bremsstrahlung radiation (1 part in 10$^{5}$). It may also be possible that a more significant amount is used to force plasma down into the chromosphere and to power chromospheric emission \citep{Woods04, Allred05}. However, despite the approximations made by EBTEL, such as the homogenous nature of the loop or the disregard for the location of energy deposition, the temperature and emission measure curves reproduce observations very well. It is computationally efficient, running a complete simulation in a matter of seconds. 

This paper has established a method that will be applied to the analysis of future events. For this case, the data was manually compared to theoretical model. However, the fitting of the parameters together with model comparison techniques is currently being investigated using a Bayesian technique for simulating values from the posterior distributions of the parameters \citep{Adamakis08}. The purpose of this analysis is to statistically optimise the model parameters within boundaries set by observations. This approach will be used when comparing theoretical models to future data sets. The authors intend to carry out an investigation of flare hydrodynamics using the improved cadence and extensive spectral range of the Extreme ultraviolet Imaging Spectrometer (EIS) on board Hinode. Combining these data with RHESSI spectral fits will vastly improve observations and allow for even more accurate modelling. 

\begin{acknowledgements}
CLR is supported by an ESA/Prodex grant administered by Enterprise Ireland. ROM would like to thank the NASA Postdoctoral Program for the Fellowship award to conduct research at the NASA Goddard Space Flight Center. The work of JAK is also supported by NASA. We would like to thank Brian Dennis, Dominic Zarro and the RHESSI team at NASA Goddard Space Flight Center for their advice and continued support. We would also like to thank the referee for their constructive advice in improving the overall quality of this paper. 

\end{acknowledgements}

\bibliographystyle{aa}

\end{document}